\documentclass[twocolumn,preprintnumbers,amsmath,amssymb]{revtex4}
\usepackage{graphicx}
\usepackage{dcolumn}
\usepackage{bm}
\usepackage{hyperref} 
\usepackage{multirow} 
\usepackage{soul}
\newcommand\beq{\begin{equation}}
\newcommand\eeq{\end{equation}}
\newcommand\bea{\begin{eqnarray}}
\newcommand\eea{\end{eqnarray}}
\usepackage{wrapfig}

\begin{document}

\title{A Density Matrix Renormalization Group 
Method Study of Optical Properties of Porphines and Metalloporphines\\}

\author{\bf Manoranjan Kumar$^{1,2}$, Y. Anusooya Pati$^1$, and S. Ramasesha$^1$}

\affiliation{\it $ {\rm ^1Solid}$ State and Structural Chemistry Unit, 
Indian Institute of Science, Bangalore 560012, India,\\
\it {\rm  $ ^2 Department$ } of Chemistry, Princeton University, Princeton, New Jersey 08544, USA \\}

\date{\today}

\begin{abstract}
The symmetrized Density-Matrix-Renormalization-Group (DMRG) method is used to study   
linear and nonlinear optical properties of Free base porphine and metallo-porphine. 
Long-range interacting model, namely, Pariser-Parr-Pople (PPP) model 
is employed to capture the quantum many body effect in these systems. 
The non-linear optical coefficients are computed within correction vector method. The
computed singlet and triplet low-lying excited state energies and their charge 
densities are in excellent agreement with experimental as well as many other 
theoretical results. The rearrangement of the charge density at carbon  and nitrogen 
sites, on excitation, is discussed. From our bond order calculation, we conclude that 
porphine is well described by the 18-annulenic structure in the ground state and 
the molecule  expands upon excitation. We have modelled the regular metalloporphine  
by taking an effective electric field due to the metal ion and computed the 
excitation spectrum. Metalloporphines have $D_{4h}$ symmetry and hence have more 
degenerate excited states. 
The ground state of Metalloporphines show 20-annulenic structure, as the charge 
on the metal ion increases. The linear polarizability seems to increase with the 
charge initially and then saturates. The same trend is observed in third order 
polarizability coefficients.

\end{abstract}

\maketitle

\section{Introduction}
 The electronic structure of porphyrin and metalloporphyrins has been
the subject of extensive study both because of their interesting optical properties
and chemical activity \cite{susli1}.
Metalloporphyrins are biologically important molecules, particularly those containing 
Fe (in Haemoglobin) and Mg (in chlorophyll) ions. The carbon atoms that form the 
conjugated back-bone undergo easy substitution leading to a large class of 
porphyrin system with interesting electronic and chemical properties. The 
$\pi-$conjugation present in these molecules gives rise to large 
nonlinear optic (NLO) responses, which can be tuned by
peripheral substitution. Porphyrins are potentially useful in applications 
such as 3-D optical memory devices \cite{3Dopt1} and optical power limiting \cite{optpowerl1}. 
Solid porphyrins are porous and could find use as molecular
sieves and shape selective catalyst \cite{susl21}. Porphyrins are
chemically and photochemically stable and are being used in photo-dynamic 
therapy \cite{photodyn1} and antiviral therapy. Oxo-Vanadium(IV)
porphyrins are studied for their anti-HIV properties \cite{yang1}. Efforts are 
being made to fabricate solar cells with porphyrin based materials \cite{solar1}, since 
porphyrin framework is found in chlorophylls. Closely related to porphyrins are the
phthalocyanines (Pc), which have a central porphyrinic core. The Pc system 
has attracted a great deal of attention in recent years for both organic electronic and 
spintronic applications \cite{pc1ref,pc2ref,pc3ref,pc4ref}. They have the interesting property 
of being adsorbed on metal surfaces in a flat orientation. Scanning tunneling 
microscopic studies have been able to measure spin and charge densities at various 
positions in the molecule.  

 Most of the important properties of porphyrins are dictated by their electronic 
structure. Hence, electronic structure of porphyrins and metalloporphyrins 
have been studied by many groups over several decades. Early studies by Goutermann 
employed a four orbital model to explain the nature of the $Q$ band (the 
low-frequency weak absorption band) and the Soret ($B$) band (the high frequency intense 
absorption band) observed in porphyrins \cite{gout}. Weiss {\it et al.} 
 used the mean field results of Pariser-Parr-Pople (PPP) model Hamiltonian 
to explain the nature of $Q$ and $B$ bands of porphyrins \cite{weiss}. 
Christoffersen {\it et al.} have carried out {\it ab initio} calculation to study 
the electronic spectra of porphyrins, metalloporphyrins and other substituted porphyrins \cite{christ}. Zerner
{\it et al.} showed that the Soret band  is degenerate with parallel
polarization with respect to $Q$ band, using random phase approximation within
 the intermediate neglect of differential overlap (INDO) method \cite{zern}. 
Using density functional theory, Ruth {\it et al.} have
studied the singlet and triplet spectra of Zn-porphyrin and related compounds
and showed that the $B$ band consists of multiple absorption bands \cite{ngub}.

Several groups have computed the NLO properties of modified porphyrins. 
Priyadarshy {\it et al.} \cite{priya} obtained the first
hyperpolarizability of porphyrin bridged donor-acceptor molecules, wherein, the
donor and acceptor molecules are attached at the $\rm meso-$position. Albert 
{\it et al.} \cite{albert} have shown that large $\rm 2^{nd}$ order NLO response 
coefficients can be obtained by attaching a donor group at $\beta$ position
 and an acceptor group at $\rm meso-$position; this is expected to increase 
the difference in the dipole moment between
ground and excited states, leading to large first order polarizability.
Shirk {\it et al.} have studied the  optical limiting properties of lead
phthalocyanine \cite{shirk}. Two-photon absorption (TPA) cross section of
porphyrin and Zn-porphyrins have been studied by a three state
model by Zhou {\it et al.} \cite{zhou}. Effect of donor-acceptor strengths on 
TPA cross section of aggregates of asymmetric Zn-porphyrins have been
studied by Ray {\it et al.} \cite{paresh}.

Porphyrin molecule is a fairly large system and hence is not amenable to accurate 
ab-initio studies. However, most of the interesting properties of porphyrin are 
associated with the $\pi-$system. Thus the electronic properties of porphyrins 
can be modeled by employing the well known PPP model for the $\pi-$electrons. 
The Porphyrinic $\pi-$system consists of 24 orbitals in conjugation occupied by 26 
electrons.  Even though the Fock space of the system is finite 
($4^{24}=2^{48} \sim  2.8 \times 10^{14}$), the large Hilbert space
of porphine, prohibits doing an exact quantum  many body calculation. Specializing 
to desired total  spin and occupancy still leaves the 
Hilbert space dimension to be very large ($9.27 \times 10^{11}$). 
Approximate techniques such as restricted configuration interaction (CI) 
for a few low-lying states is not very reliable, since it relies on cancellation 
of errors, as in single CI. Thus it is important to bring to bear novel and accurate 
methods for important low-lying states as well as NLO response coefficient of 
the systems. 

    In this paper, we have employed the Density Matrix Renormalization Group 
(DMRG) method to carry out reliable model many-body calculations of the ground 
and excited state properties. In the next section we give a brief introduction 
to the DMRG method. In section III, we discuss the model Hamiltonian and 
implementation of the DMRG method to porphines. In section IV, we present our 
results on the properties of the low-lying states of porphine. In section V, we
discuss the dynamic nonlinear optic coefficient computed for the system.

\section{Introduction to the DMRG method} The DMRG algorithm involves 
efficient and accurate ways of truncating the insignificant degrees of freedom 
from the Fock space of the system. The DMRG method proceeds as follows. Given a 
model many body Hamiltonian $\hat H $  which we wish to solve for a system with total 
number of sites $N$. We 
start with a small lattice of $2L$ sites and obtain the 
Hamiltonian matrix of this lattice in the real space basis. We obtain by 
standard methods, the lowest or the desired eigenstates of the 
Hamiltonian matrix. From the eigenvector we construct the  
density matrix $\rho$, of the subsystems with L sites by integrating out 
the states of the remaining L sites. The matrix elements of the density matrix, 
$\rho^{k,L}_{s,s'}$ of the L sites block in the $k^{th}$ eigenstate is given by,   
\begin{eqnarray}
\rho^{k,L}_{s,s'}=\sum_{e} C_{s,e}^{k} C_{s',e}^{k} 
\end{eqnarray}
where the $k^{th}$ eigenstate of the 2L sites problem is expressed as
\begin{eqnarray}
\psi_{k,2L}= \sum_{s}\sum_{e} C_{s,e}^{k}|s \rangle |e\rangle,
\end{eqnarray}
 with $|s \rangle$ and $|e \rangle$ being the Fock space basis states of 
the system and environment blocks respectively and $C_{s,e}^{k}$ are the 
associated coefficients. The density matrix $\rho^{k,L}$ is diagonalized 
and eigenvectors corresponding to $m$ dominant eigenvalues of $\rho^{k,L}$ 
are used to span the Fock space of the system block. If the order of the matrix
 $\rho^{k,L}$ is $M\times M$, then the $m\times M$ matrix $O $ 
formed from $m$ eigenvectors as columns of the $O$ is used for  
transforming all the operators of the system block, L, from the full 
Fock space to the truncated density matrix eigenvector basis. 
Thus if A is an $M\times M$ matrix  of an operator $\hat A$ defined 
over the $M$ dimensional Fock space of the system block, 
then $\tilde A=OAO^{\dagger}$ gives the renormalized $m\times m$ 
matrix  representation of $\hat A$. By interchanging 
the environment and system blocks, we also obtain renormalized matrix 
representation of all the operators in the former environment block. 

We now add two new sites to the $2L$ size system at a convenient 
part of the system, usually in the middle. Using as basis functions, 
the direct product of the density matrix eigenvectors of the two 
blocks of the $2L$ system and the Fock states of the new sites, we can 
obtain a matrix representation of the $2L+2$ site Hamiltonian. From the 
desired eigenstate $\psi_{k, 2L+2}$ we obtain the density matrix $\rho^{k,L+1}$ 
for the two blocks of size $(L+1)$ of the augmented (2L+2) site system 
and proceed as before to obtain all the renormalized operators on the 
$(L+1)$ sites and iterate until the desired system size is reached. This 
algorithm of constructing the system is called the infinite DMRG algorithm. 
In this algorithm, from the eigenvectors at any system size, 
we can also obtain expectation values of any operator of interest. 
If the operator consists of product of site operators of different 
sites in the same block, then we need to carry through matrix 
representation of the product by renormalizing it at each step 
and use the renormalized matrix to compute the desired expectation value. 
If on the other hand the operator is a product of site operators of 
different sites on different blocks, we can use the renormalized 
matrix representation of the individual site operators to compute 
the expectation values. It is possible to target more than one low-lying 
state simultaneously in a DMRG procedure. This is carried out by computing 
the density matrix  for each desired states separately and using 
the dominant eigenvectors of an appropriately averaged density matrix of these states. 
The choice of common DMRG basis for different states permits computing 
matrix elements such as transition dipoles between states. To access 
higher excited states of interest such as the $1 ^1B_u$ states in 
polyenes, we need to employ symmetrized DMRG procedure and compute 
the desired excited state as a low-lying state in an appropriate 
symmetry subspace. For example, we could access a triplet 
state as the lowest energy excitation in a space of odd parity, other parity
symmetry corresponds to the invariance of the Hamiltonian when the spin space is 
rotated by $\pi$ around the Y-axis.

For molecular systems the infinite DMRG technique is not 
sufficiently accurate since the density matrix of the system block does not 
correspond to the density matrix of the target superblock. This can can be 
remedied by employing the finite DMRG algorithm. In this technique the infinite 
algorithm is employed till the desired superblock is attained. After this point, 
the system block is augmented by one site and the Hamiltonian matrix of the superblock 
is set up using the larger system block, two new site and an environment block with 
one fewer site generated in the infinite algorithm. From  the desired eigenstate of the 
Hamiltonian matrix, the system block is again augmented by one site and the whole 
process is repeated until the environment block reduces to a single site. From 
here on, the environment block and system block are interchanged and the process 
continued until we reach the size of environment and system block
achieved at the end of the infinite DMRG  algorithm. This process of progressively 
increasing the size of the system block and decreasing the size of the environment 
block, back and forth is called finite DMRG algorithm \cite{white2}. 

The DMRG technique has been shown to be highly accurate for quasi-one dimensional
systems and has been widely employed in the study of the strongly correlated model
Hamiltonians such as the Hubbard models \cite{hub1}, the PPP 
models \cite{ppp1,ppp2,ppp3,pppmano} and the Heisenberg spin models 
\cite{spin1,spin2,spin3,revdmrg}. The PPP model has long 
range interactions and topologically the model can be viewed as a higher 
dimensional model.  However, the interaction terms in the PPP model are diagonal 
in the real space and are hence well accounted for in the DMRG method.

  In this work, we report our studies on free base porphine (FBP) and 
metallo-porphines obtained from DMRG studies of their electronic structure. 
Porphine is modeled by PPP model with Ohno parametrization for the long-range 
interactions.  We have studied the optical excitation spectrum by using 
symmetrized DMRG method \cite{sr2}. We characterize optically excited states 
by studying various properties of the states, such as, charge densities 
and bond orders. We have also studied both linear and nonlinear optical 
properties of the system.
\begin{figure}
\begin{center}
\hspace*{0cm}{\includegraphics[width=8.0cm,height=8.0cm]{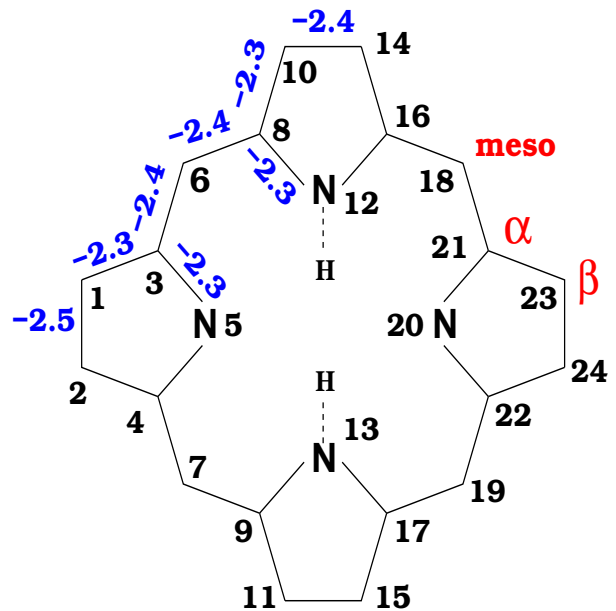}} \\
\caption{General structure of FBP. Sites are numbered in bold and 
transfer integrals (in eV) are given in blue colour. Labelling of the sites
meso, $\alpha$ and $\beta$ are also shown in the figure. The system has 
$D_2$ symmetry  and all other bond transfers are given by symmetry.
}
\label{porfig1}
\end{center}
\end{figure}

\section{Methodology}
Porphine contains 4 pyrrolic rings connected by 4 $\rm CH$ groups as shown in 
Fig. \ref{porfig1}. Nitrogen in the pentagons are $sp^2$ hybridized. There are 
two types of nitrogen atoms in the conjugation, one is the aza nitrogen and the other 
is the pyrrole like nitrogen.
Lone-pair of electrons on N are in the plane of the molecule in  aza-ring systems and 
do not participate in conjugation, whereas, they are perpendicular to the plane of 
the ring and  are involved in $\pi$-conjugation, in pyrrole rings. The singly 
occupied $p_z$ orbital of the aza nitrogens are involved in $\pi$-conjugation. 
Each carbon atom in the conjugation contributes one electron to $\pi$-system. 
Thus, in free base porphine, 26 $\pi-$electrons and 24 orbitals are involved in 
conjugation. It has been established that PPP model is  well suited 
for describing the electronic properties of conjugated organic molecules. The PPP model 
Hamiltonian is given by 
\begin{eqnarray}
\hat H_1&=& \sum_{i,\sigma}\epsilon_i \hat a^\dagger_{i,\sigma} \hat a_{i,\sigma}+
\sum_{<i,j>,\sigma} t_{ij} \hat a_{i,\sigma}^ {\dagger}\hat a_{j,\sigma}+h.c  \\   \nonumber
\hat H_2&=& \sum_{i} U_{i} \hat n_{i,\sigma} \hat n_{i,\sigma^{'}}\\ \nonumber
\hat H_3&=& \sum_{i>j} {V_{i,j}(\hat n_i-z_i)(\hat n_j-z_j)}\\ \nonumber
\hat H_{PPP}&=&\hat H_1+\hat H_2+\hat H_3
\end{eqnarray}

Here $H_1$ corresponds to the one-electron part of the Hamiltonian with  
the orbital energy, $\epsilon$, of $p_z$ orbital and  $t_{ij}$, the transfer integral 
between the orbitals $i$ and $j$ on the bonded pair of atoms in conjugation. We discuss 
the parametrization 
involving N-orbitals later. We have used experimental geometry for our calculations \cite{webb}. The
transfer integrals, $t_{ij}$ are calculated using a linear extrapolation. i.e.
\begin{equation}
t_{{\rm ij}} = t_0 (1-\frac {(\delta r)_{ij}}{r_{ij}})
\label{eq22}
\end{equation}
$t_0$ is the transfer integral ($-2.4 eV$) corresponding to  $\rm {C-C}$ bond length
of $1.397  $ \AA ~~in benzene and $(\delta r)_{ij}$ is (${r_{ij}-1.397}$) \AA ~~where 
$r_{ij}$ is the length of the bond between atoms i and j. 

The term $H_2$ in the PPP Hamiltonian incorporates on-site electron-electron repulsion energy and 
this corresponds to Hubbard interaction. The Hubbard parameter  
$U_{i}$ is the energy cost for creating double occupancy of the orbital $i$. 
For carbon we have taken the standard value of $11.26$ eV. The terms in $H_3$
correspond to inter-site coulombic interaction between
electrons in orbitals i and j, and we introduce local chemical potential
 $z_i$ which is the occupancy of the orbital $i$ that leaves the $i^{th}$ site 
electronically neutral. The value of $z$ for C is always 1 while that of nitrogen
depends upon the occupancy of the orbital involved in conjugation, for the 
aza-nitrogen which contributes one electron to the $\pi-$conjugation, $z$
 is 1, while for the pyrrole nitrogen which contributes two electrons to the 
$\pi-$conjugation, $z$ is 2; $V_{i,j}$ is calculated using Ohno parametrization \cite{ohno2}.

We have parametrized $\epsilon$ and $U$ for nitrogen atom to fit the experimental 
excitation gaps and corresponding transition dipole moments for 
singlets in free base porphine. The parameters which give the best fit
are as follows: for aza nitrogens, $\epsilon_{N_{5}}=~\epsilon_{N_{20}}=-3.20~ eV$ and
$U_{N_{5}}=~U_{N_{20}} = 12.34~ eV$; for pyrrole nitrogens,  $\epsilon_{N_{12}}=~\epsilon_{N_{13}}=-14.0~ eV$ and
$U_{N_{12}}=~U_{N_{13}}~=15.00~ eV$. The value of $t_{CN}$ is taken to be $-2.3~eV$ and
$-2.4 eV $ for ${\rm C-N}$ single and double bonds respectively. Other transfer integrals
 calculated from Eq. \ref{eq22} are shown in Fig. \ref{porfig1}. The nitrogen parameters 
reflect the more compact $2p$ orbitals of nitrogen compared to $2p$ orbitals of carbon. 

\begin{figure}
\begin{center}
\hspace*{0cm}{\includegraphics[width=9cm, height=9.0cm,angle=-0]{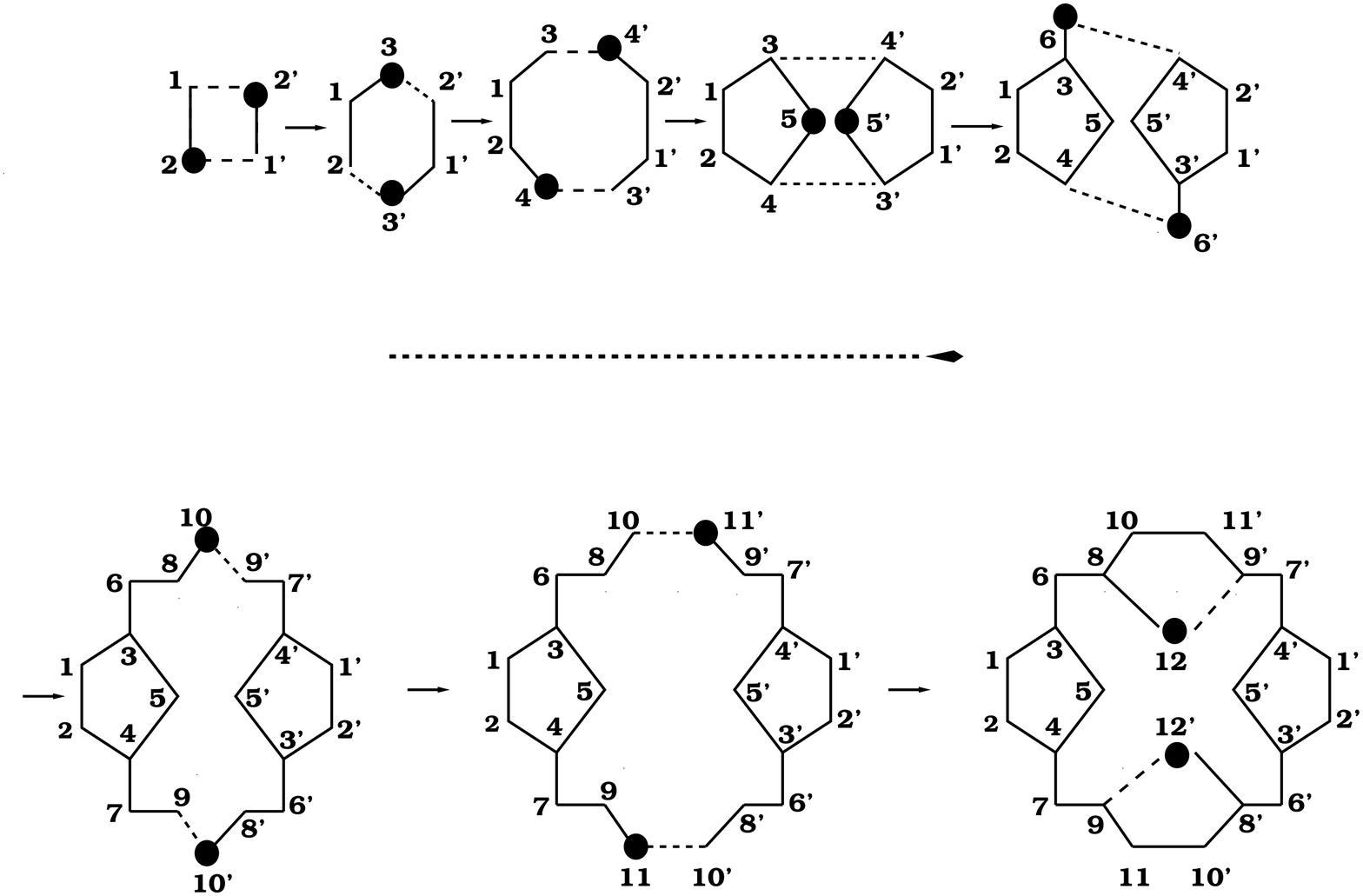}} \\
\caption{A highly accurate scheme for building the porphine structure for DMRG
calculations. At every step of the DMRG algorithm, we add two new sites shown 
by filled circle. Positive integers correspond to the sites of the left 
block and negative integers to the sites of the right block.}
\label{porfig2}
\end{center}
\end{figure}

We have employed the DMRG methods to 
study the ground state (gs) and low-lying states of porphine. The topology of the 
transfer terms in porphines is not strictly one dimensional. The porphine system 
can be built up from a 4-site ring, by adding two sites at a time, in many 
different ways. The scheme shown in Fig. \ref{porfig2} provides a highly 
accurate scheme for building up the porphine framework, besides retaining 
the symmetries of porphine at every stage of the DMRG implementation. We have also carried out 
finite DMRG sweeps to improve the accuracy. The accuracy of the DMRG scheme is bench-marked 
against  H\"uckel  model results for the system. In the DMRG studies, we have retained 
210 density matrix eigenvectors (DMEVs) and carried out two finite DMRG sweeps. The 
order of the Hamiltonian matrix is about $45,000$ after using two symmetries, namely, 
spin invariance and $\rm C_2$ symmetry. The matrix is sparse and symmetric. We obtain a 
few low-lying states of the matrix using Davidson algorithm. Various properties 
such as charge and spin densities and b.o.s in different eigenstates can be 
computed since the matrices of the creation operators, the occupation number
 operators of all the sites and the b.o operators of all the bonds are 
renormalized at every step. 

In the molecular systems with strong electron correlations, it 
is difficult to identify the desired excited states such as two 
photon states or optically allowed excited states, unless
symmetry is exploited. The Hamiltonian is spin conserving, however, 
exploiting the conservation of both $S^2$ and $S^z$ 
($S$ corresponds to total spin ) within a DMRG scheme is 
difficult. Instead we have used spin invariance symmetry and $S^z$ conservation, 
to reduce the dimensionalities of the spaces. Spin invariance symmetry
corresponds to invariance of the Hamiltonian when all the 
spins are rotated by $\pi$ around the y-axis, within the 
$M_s=0$ ($M_s$ is eigenvalue of $S^z$ operator) subspace. The spin invariance
operator divides the basis into two subspaces, one of even 
total spin and another of odd total spin. The spatial symmetry 
we have employed is the $C_2$ symmetry, even though the molecule has $D_2$ symmetry.
The $C_2$ axis chosen by us passes through 
the center of the molecule and is perpendicular to molecular plane. 
The spin invariance symmetry commutes with the $C_2$ symmetry and the 
resulting group has four one dimensional representations. 
Optical transitions are from states which are even under $C_2$ 
symmetry to those that are odd, and retain total spin. The 
triplet (S=1) state is targeted as the lowest state in the 
$M_s=1$ sector. Since we are dealing with a nonzero $M_s$ 
subspace, the spin invariance symmetry does not exist in this case and 
we only employ the spatial symmetry. Besides, for computing 
properties such as transition dipoles and dynamic linear and 
nonlinear polarizabilities, we need to express  states in the different 
symmetry subspace by using the same DMRG basis states. In order to retain 
proper balance of various states, we construct an average density matrix, 
averaged over density matrices of several low-lying states in each 
subspace and employ dominant eigenvectors of the averaged density matrix. 
The dynamic nonlinear optics response coefficient are computed using 
correction vector (CV) technique \cite{cplsr}.

Conventional frequency dependent linear and nonlinear optical
coefficient calculations rely on the knowledge of transition 
dipole moments between the gs and all excited states. 
For exact calculation of NLO properties, the sum over states 
(SOS) method is most widely used. Generally, 
between 30 to 100 excited states, computed within   
a restricted CI scheme are used to obtain NLO coefficients. Soos 
and Ramasesha \cite{soos_rama} introduced the CV method for calculating 
NLO coefficients within an exact diagonalization scheme 
which incorporates the contribution of all higher excited states, without explicitly 
obtaining the excited state eigenfunctions. In this method, 
a correction vector (CV) is obtained using the gs wavefunction and
the Hamiltonian matrix. CV method involves computation of the first 
and second order correction vectors $\phi^{(1)}_i (\omega_1)$ 
and $\phi^{(2)} _{ij} ( \omega_1,\omega_2)$, which are defined as,

\begin{center}
\begin{equation}
({\mathbf H}-E_G+\hbar \omega_1 + i \Gamma) |\phi^{(1)}_i (\omega_1)> =
\tilde \mu_i|G> 
\end{equation}
\end{center}

\begin{center}
\begin{equation}
({\mathbf H}-E_G+\hbar \omega_2 + i \Gamma) |\phi^{(2)}_{ij}
(\omega_1,\omega_2)> 
= \tilde \mu_j |\phi^{(1)}_i (\omega_1)>
\end{equation}
\end{center}

\noindent
where $\mathbf H$ is the Hamiltonian matrix in the chosen many-body basis,
$E_G$ is the gs energy, $\omega_1,~\omega_2$ are the excitation 
frequencies and $\tilde\mu_i$ is the $i^{th}$ component of the dipole displacement
operator, ($\tilde{\mu_i}$ =$ \hat{\mu_i} - <G|\hat {\mu_i}|G>$), and $\Gamma$
is the average lifetime of the excited states. It can be shown that
$\phi^{(1)}_i (\omega_1)$ and $\phi^{(2)}_{ij} ( \omega_1,\omega_2)$ when expressed
in the basis of the eigenstates of the Hamiltonian ${|R>}$ ~are given by,

\begin{equation}
 |\phi^{(1)}_i (\omega_1)> =  \sum_R \frac {<R | \tilde \mu_i | G>} 
{E_R - E_G + \hbar \omega_1 + i \Gamma} |R> 
\end{equation}

\begin{eqnarray}
\lefteqn
{|\phi^{(2)}_{ij}( \omega_1,\omega_2)>=} \nonumber \\
 &&\sum_S\sum_R \frac {<S|\tilde\mu_j|R><R|\tilde\mu_i|G>}
{(E_R -E_G+\hbar\omega_1+i\Gamma) (E_S -E_G+\hbar\omega_2+i\Gamma)} \nonumber \\ 
& &|S>  
\end{eqnarray}

\noindent
Therefore  $|\phi^{(1)}_i (\omega_1)>$ and $|\phi^{(2)}_{ij} ( \omega_1,
\omega_2)>$ can be readily used to compute linear and nonlinear frequency
dependent polarizabilities. The third order NLO coefficients corresponding to 
sum frequency generation, in terms of these correction vectors, are given by,

\begin{eqnarray}
\lefteqn
{\gamma_{ijkl}(-\omega_{\sigma}; \omega_1,\omega_2,\omega_3,)=} \nonumber \\
& &\hat P_{ijkl}( <\phi^{(1)}_i(-\omega_\sigma)| \hat{\mu_j}| 
\phi^{(2)}_{kl} (-\omega_1-\omega_2, -\omega_1)>) 
\end{eqnarray}

\noindent
where  $\hat P_{ijkl}$ generate all permutations of 
$(-\omega_\sigma,i)$, $(\omega_1,j)$, $(\omega_2,k)$,
$(\omega_3,l)$ leading to 24 terms for
$\gamma$ ~(with $\omega_\sigma$ = $\omega_1+\omega_2+\omega_3$). The linear 
polarizability components are given by
\begin{eqnarray}
\alpha_{ij}(\omega)& =&(<\phi^{(1)}_i(\omega)|\tilde{\mu_j}|G> +
                         < \phi^{(1)}_i(-\omega)|\tilde{\mu_j}|G>) 
\end{eqnarray}
The $\phi_i$s are exact within the Hilbert space chosen for the Hamiltonian. 
The linear algebraic equations that can be solved efficiently by a 
small matrix algorithm developed by Ramasesha \cite{sralgo}.  

Ramasesha {\it et al.} incorporated the CV technique in the DMRG method 
and have shown it to be robust \cite{ramaswapn}. In the CV-DMRG 
procedure, the average density matrix is constructed from a weighted 
average of the gs and the density matrix constructed from the 
correction vector. We will concentrate on the spatially 
averaged values of first and third order optical response. The 
expression for these are given by,
\begin{eqnarray}
\alpha_{av} & = & \sum_{i=1}^3 \frac {1}{3} \alpha_{ii} \\
\gamma_{av} & =&  \sum_{i,j=1} ^3 \frac {1} {15}  ( 2 \gamma_{iijj} + \gamma_{ijji})
\end{eqnarray}
where $\{i \}$s are Cartesian indices (x,y,z). 

\noindent
\begin{table}
\begin{center}
\caption {Comparison of excitation gaps and transition dipole moments 
between DMRG and exact calculations, in the non-interacting limit. Gaps are given
in eV and transition dipoles are in a.u.}
~\\
\begin{tabular}{|ccc|ccc|ccc|c|} \hline
 \multicolumn{6}{|c}{DMRG} & \multicolumn{3}{|c|}{Exact} \\\cline{1-6}
\multicolumn{3}{|c}{$m=180$ } & \multicolumn{3}{|c}{$m=210$} & \multicolumn{3}{|c|}{} \\\hline
Gap & $|\mu_x|$  & $|\mu_y|$ &  Gap& $|\mu_x|$ & $|\mu_y|$ &  Gap & $|\mu_x|$ & $|\mu_y|$ \\\hline
1.67 & 0.17 & 2.74& 1.65 & 0.22 &  2.73 &   1.50 & 0.44 &2.76 \\
1.82 &1.07 & 0.09& 1.78 & 1.47 &  0.32 &   1.56 & 2.37 & 0.52 \\ 
1.96 & 0.45 & 2.38 &  1.95 & 0.65 &  2.36 &   1.80  & 2.25 & 0.44 \\
2.05 &2.90 & 0.24 &  2.03&  2.72 & 0.51  & 1.87  &0.42 & 2.35 \\\hline
\end{tabular}
\label{portb1}
\end{center}
\end{table}
We have compared our DMRG results for the non-interacting Hamiltonian
with exact H\"uckel MO results for porphine. 
 In the above calculations, at each iteration, the density matrix is
constructed  as the weighted average of the density matrix of five 
lowest eigenstates. Table \ref{portb1}. compares the optical gaps and
transition dipole moments for two different DMRG cut-offs, $m$, from finite DMRG 
calculations with exact results. We observe that $m= 210$ gives
quite accurate results and keeping higher $m$ increases the demand on  computational
resources with only marginal improvement in the results. DMRG results slightly
overestimate the higher excited states ($\approx$ 0.15 - 0.20 eV). Due to
near degeneracy in the excitation levels, the components of transition dipole 
moments do not agree very well with  H\"uckel MO calculations although the magnitudes
are in good agreement and the qualitative trend is
maintained. All the studies that we report in this paper are based 
on retaining 210 density matrix eigenvectors and two sweeps of the finite algorithm.

\section{Results and Discussion}
This section is divided into three subsections. In the next subsection, 
we discuss results of our study on FBP followed by the metallo-porphine. In the last section, 
linear and nonlinear optic properties are discussed. 

\subsection{Free base porphine}     
We have obtained six lowest lying states in A and B subspaces for singlets 
as well as triplets of FBP using symmetrized DMRG method. In Table \ref{portb2} we 
present the energies of singlets and in Table \ref{portb3} that of triplets. The ground 
state energy is set to zero, so the energies reported are the excitation gaps from the gs. 
We have also presented experimental gaps and absorption intensities, wherever available.
 DMRG results correspond to isolated molecule calculations in the gs geometry. 
The experimental optical gaps are usually red shifted by about 
0.5 eV from the gas phase values. The first peak observed in the optical spectra 
is at 1.98 eV while that calculated is at 1.66 eV. If we include the red shift of the theoretical gap,
the agreement is off by about 0.8 eV. This suggests that the first peak in the spectra may be due to 
$n-\pi^*$ transition, and our model being a purely $\pi$-electron model cannot account 
for such a state. The 1.66 eV transition in the solid state would be shifted to about 1 eV 
corresponding to approximately 1200 nm, and will not be observed in an optical spectra. 
The 2.86 eV absorption to the $2^1B$ state is X-polarized (see Table \ref{portb2}) 
while the nearly degenerate $3^1B$ and $4^1B$ states at 3.89 eV and 3.91 eV 
peaks have X and Y polarizations respectively. It should be noted that the 
low-temperature spectra of the free base porphine shows a split in the $B$ band 
with equal intensity and separated by 0.03 eV \cite{rimi}. Other theoretical studies 
also show nearly degenerate levels at this energy. The 4.18 eV band ($5^1B$) 
corresponds to the $N$ band while the 4.83 eV  absorption to $6^1B$ corresponds to 
the weak $L$ band transitions observed in FBP.
\begin{table}
\begin{center}
\caption{Excitation gaps of six low-lying excited states and corresponding
transition dipole moments compared with the experimental results for optically 
allowed states. Calculated oscillator strengths are given in parenthesis. Experimental 
results are from reference \cite{dolphin}. The experimental oscillator strengths are 
normalized with respect to most intense absorption. Sum of calculated 
intensity ($\mu_x^2$+$\mu_y^2$) for the state 3 and 4 is taken to be unity as they 
are nearly degenerate and all others are normalized with respect to it. 
Transition dipole between gs and all excited $A$ states  strictly vanishes by
 symmetry. }  
~\\
\begin{tabular}{|ccccc|} \hline
State  & Optical gap & $|\mu_x|$ & $|\mu_y|$ & Observed  \\
label &  (in eV)  &  (in a.u.) &   (in a.u.) & value \\ \hline
$ 1^1B$ &  1.66 (${\bf 0.16}$) &  0.05 &  0.30 & 1.98 ${\bf (0.01)}$  \\
$ 2^1A$ &  2.94        &   -    & -       & -  \\
$ 2^1B$ &  2.86 ${\bf (0.26)}$ &  0.39 &  0.05 & 2.42 ${\bf (0.05)}$ \\
$ 3^1A$ &  3.40        &   -    & -     & -   \\
$ 4^1A$ &  3.53        &  -      &-      & -   \\
$ 3^1B$ &  3.89 ${\bf(1.00)}$ &   0.48 & 0.06 & 3.33 ${\bf (1.00)}$  \\
$ 4^1B$ &  3.91 ${\bf (1.00)}$ &   0.10 & 0.59 &             \\
$ 5^1A$ &  4.01        &   -     &-      & -  \\
$ 5^1B$ &  4.18 ${\bf (0.21)}$ &  0.04 &  0.35 & 3.65 ${\bf (0.87)}$ \\
$ 6^1A$ &  4.33        &   -     & -      & -  \\
$ 6^1B$ &  4.83 ${\bf (0.01)}$ &  0.00 &  0.07 & 4.25 ${\bf (0.87)}$   \\ \hline
\end{tabular}
\label{portb2}
\end{center}
\end{table}

\begin{table}
\begin{center}
\caption{Excitation gaps of low-lying excited states and transition dipole 
moments in triplet  manifolds. Note that all the excitations are from 
$B$ to $A$ space. Experimental results are from references \cite{trexpa} 
and \cite{trexpb}.}
~\\
\begin{tabular}{|cccccc|} \hline
State &  Triplet energies & T-T gap & $\mu_x$ & $\mu_y$ & Observed  \\
label & (in eV) &    (in eV)  & (in a.u.) &   (in a.u.) & value \\ \hline
 $1^3 B $ &  1.42   & - &-     & - &  -      \\
 
 $2^3 B $ &  2.45   &  1.03  &-     & - &  -      \\
 $3^3 B $ &  2.93   &  1.51  &-     & - &  -     \\
 $1^3 A $ &  3.09   &  1.67  &0.00  &0.18 & 1.58   \\
 $2^3 A $ &  3.51   &  2.09  &0.03  &0.02 & 1.65   \\
 $4^3 B $ &  3.55   &  2.13  &- & - &  -     \\
 $3^3 A $ &  3.55   &  2.13  &0.05  &0.08 & -  \\
 $4^3 A $ &  4.26   &  2.84  &0.00  &0.29 & 2.82 / 2.96  \\
 $5^3 B $ &  4.28   &  2.86  &- & - &  -     \\
 $5^3 A $ &  4.83   &  3.41  &0.02  &0.12 &  3.23  \\\hline
\end{tabular}
\label{portb3}
\end{center}
\end{table}

In  Table \ref{portb3}, we give energies of several low-lying triplet states. 
The lowest singlet-triplet gap in the system is 1.42 eV from our calculations, 
 experimental value is 1.58 eV, which is in fair agreement. The lowest 
triplet-triplet absorption, from our calculations is found at  1.67 eV, while 
experimentally two very close triplet-triplet absorption peaks are found at 
1.58 eV and 1.65 eV \cite{sapunov,trexpb}. 
Our calculations show weak absorption peaks at 2.09 eV and 2.13 eV and a much 
stronger absorption is predicted at 2.84 eV. Experimentally two absorption 
peaks are seen at 2.82 and 2.96 eV, close to the theoretical values. Theory also
predicts a triplet absorption at 3.41 eV while experimentally a triplet absorption 
is seen at 3.23 eV. Overall, we find that the DMRG results for  the 
$\pi$-electron model are quite consistent with experimental observations.

\subsubsection{Electron density and bond order analysis}
\begin{figure}
\begin{center}
\hspace*{0cm}{\includegraphics[width=9.0cm,height=9.0cm]{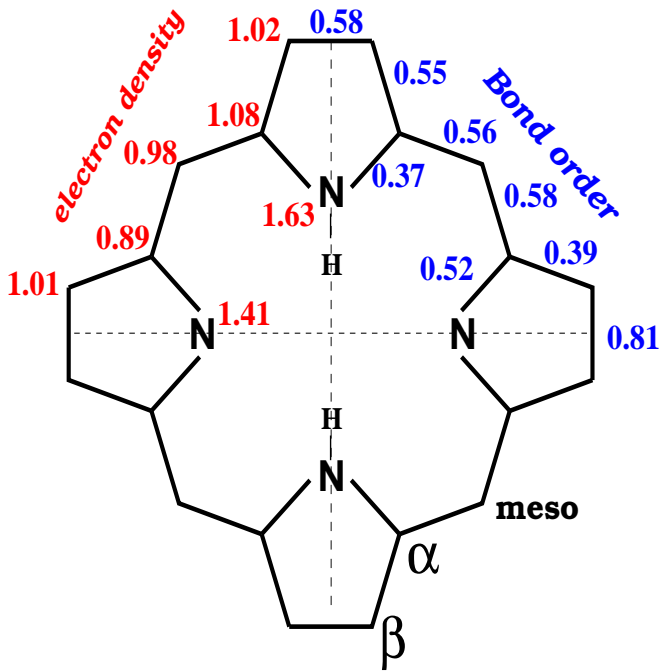}} \\
\end{center}
\caption{ Electron density and b.o. for the gs of porphine. 
Charge densities are shown on the left part of the structure while b.o.s 
are shown on the right part. The site indices and bond indices are given in
red and blue, respectively, for half of the system. The $\rm D_2$ symmetry gives value 
for remaining sites and bonds of the molecule. }
\label{porfig3}
\end{figure}

We have analyzed the eigenstates by computing charge densities 
and b.o.s in various states. In the gs, 
all $\beta$ carbon atoms have slightly more than one electron. The electron 
density at $\alpha$ carbon atoms in aza ring is $0.89$ and
in pyrrole ring it is $1.08$ and have opposite acceptor (donor) tendencies. At 
$meso-$carbon, the charge density is $0.98$. 
This is consistent with the earlier semi-empirical calculations which 
show that the $\rm C_{meso}$ is electron deficient and $C_{\beta}$ is slightly
electron rich \cite{albert}. Nitrogen in the aza ring has 1.41 electrons and in 
pyrrole ring it has 1.63 electrons. These results are in
agreement with the earlier theoretical work \cite{weiss}. The gs
charge density and  b.o. are given in Fig. \ref{porfig3}. It is interesting 
to note that although aza ring nitrogen contributes only one electron to $\pi$
-conjugation the electron density is 1.41. 

\begin{table}
 \caption{Electron density  $\rho$ for gs and difference of electron
density $\delta\rho$  for the optically allowed states w.r.t. gs. 
}
\begin{center}
\resizebox{8cm}{!}{
\begin{tabular}{|c|c|rrrrrr|} \hline
&\multicolumn{1}{c}{$\rho$} & \multicolumn{6}{|c|}{$\delta\rho$} \\\cline{2-8}
Site & & & & & & & \\
&  gs ($A_0$) & $B_1$ & $B_2$ & $B_3$ & $B_4$ & $B_5$ & $B_6$ \\\hline
$C_{\beta}$ ~(aza) &1.01 &   0.01&  0.00& -0.00& -0.03& -0.04& -0.00  \\
$C_{\beta}$ ~(aza)  &1.01 &   0.00& -0.00& -0.01& -0.02& -0.05& -0.01  \\
$C_{\alpha}$ ~(aza)  &0.89 &  -0.01&  0.01& -0.01&  0.04&  0.03&  0.02  \\
$C_{\alpha}$ ~(aza)  &0.89 &  -0.01&  0.02&  0.01&  0.05&  0.05&  0.01  \\
$N$ ~(aza)  &1.41 &  -0.02&  0.00&  0.02& -0.11& -0.08& -0.03  \\\hline
$C_{meso}$ &0.98 &   0.01&  0.00&  0.03&  0.03&  0.03&  0.01  \\
$C_{meso}$ &0.98 &   0.00& -0.00&  0.01&  0.02&  0.03&  0.02  \\\hline
$C_{\alpha}$ ~(pyr)  &1.08 &  -0.02&  0.04&  0.02& -0.03& -0.02&  0.04  \\
$C_{\alpha}$ ~(pyr)  &1.08 &  -0.01&  0.03&  0.03& -0.03&  0.01&  0.02  \\
$C_{\beta}$ ~(pyr)  &1.02 &  -0.00&  0.04& -0.01&  0.00&  0.03& -0.00  \\
$C_{\beta}$ ~(pyr)  &1.02 &   0.01&  0.04& -0.05&  0.03&  0.02&  0.03  \\
$N$ ~(pyr)  &1.63 &   0.04& -0.19& -0.05&  0.06& -0.00& -0.11  \\ \hline
\end{tabular}}
\label{portb4}
\end{center}
\end{table}
For excited states which are Y-polarized, electron density is transferred 
from nitrogen in aza ring to the neighboring carbon atoms while for the 
the X-polarized excited state the electron is transferred from the carbon atom 
to the $N$ atom. Except for the fourth and fifth excited states, for 
which there is considerable transition dipole moment, the change in 
electron density on $\rm C_{\alpha}$ and $ \rm C_{\beta}$ sites are negligible. 
In the fourth 
and fifth excited states, there is considerable electron density redistribution 
between carbon and nitrogen atoms. The electron density at meso sites
remain almost the same in all the excited states.

In pyrrole rings the opposite trend of electron transfer is observed except for
the sixth excited state. The electron is transferred to N when the polarization
is in $Y$ direction and is transferred from $N$, when it is in $X$ direction.
The magnitude of electron distribution is maximum in second excited state. In the fifth 
excited state, there is hardly any change in the electron
distribution at $N$ atom, but the charge is transferred between $ \rm C_{\alpha}$ and
$\rm C_{\beta}$ atoms. Deviation in electron density from the gs are
given in Table \ref{portb4}.
\begin{table}
\begin{center}
\caption {Bond orders of gs and difference in b.o. for six low-lying states w.r.t gs, in the 
singlet manifold are given. Symmetry space of states are given in  parentheses. }
~\\
\resizebox{9cm}{!}{
\begin{tabular}{|c|c|c|c|c|c|c|c|} \hline
Bond &  b.o & $\Delta(b.o)$  & $\Delta(b.o)$ &$\Delta(b.o)$ & $\Delta(b.o)$ & $\Delta(b.o)$ & $\Delta(b.o)$\\
&  G.S (1A)  & ($1^{st}$)(1B) & $2^{nd}$ (2A) & $3^{rd}$ (2B) & $4^{th}$ (3A)  & $5^{th}$ (4A)  & $6^{th}$ (3B)  \\\hline
$C_{\beta}-C_{\beta}$ ~(aza)  & 0.81 &    -0.01&   -0.01&   -0.03&   -0.15&   -0.16&   -0.04 \\
$C_{\beta}-C_{\alpha}$ ~(aza)   & 0.39 &     0.00&    0.01&    0.02&    0.10&    0.08&    0.03\\
$C_{\beta}-C_{\alpha}$ ~(aza)   & 0.38 &     0.00&    0.01&    0.01&    0.11&    0.08&    0.02\\
$C_{\alpha}-N$ ~(aza)   & 0.54 &    -0.01&   -0.02&   -0.03&   -0.06&   -0.07&    0.00\\
$C_{\alpha}-N$ ~(aza)   & 0.52 &    -0.01&   -0.01&    0.00&   -0.02&   -0.05&   -0.03\\ \hline
$C_{\alpha}-C_{meso}$   & 0.59 &    -0.03&   -0.01&   -0.02&   -0.10&   -0.03&   -0.01\\
$C_{\alpha}-C_{meso}$   & 0.58 &    -0.03&    0.01&    0.02&   -0.06&   -0.02&   -0.04\\
$C_{\alpha}-C_{meso}$    & 0.55 &    -0.01&   -0.02&   -0.02&    0.02&   -0.04&   -0.03\\
$C_{\alpha}-C_{meso}$ & 0.56 &    -0.01&   -0.03&   -0.06&    0.00&   -0.06&   -0.01\\\hline
$C_{\alpha}-C_{\beta}$ ~(pyr) &  0.57 &    -0.01&    0.04&    0.04&   -0.02&    0.06&    0.02\\
$C_{\alpha}-C_{\beta}$ ~(pyr)  &  0.55 &    -0.01&    0.04&    0.05&    0.00&    0.05&    0.01\\
$C_{\beta}-C_{\beta}$ ~(pyr)  &  0.58 &    -0.04&   -0.12&   -0.19&   -0.07&   -0.14&   -0.18\\
$C_{\alpha}-N$ ~(pyr) &  0.37 &    -0.02&   -0.06&   -0.04&   -0.05&   -0.06&   -0.11\\
$C_{\alpha}-N$ ~(pyr) &  0.38 &    -0.03&   -0.03&   -0.02&   -0.06&   -0.04&   -0.09\\ \hline
\end{tabular}}
\label{portb5}
\end{center}
\end{table}

\begin{figure}[htbp]
  \begin{center}
 {\includegraphics[height=6.0cm, width=8.5cm]{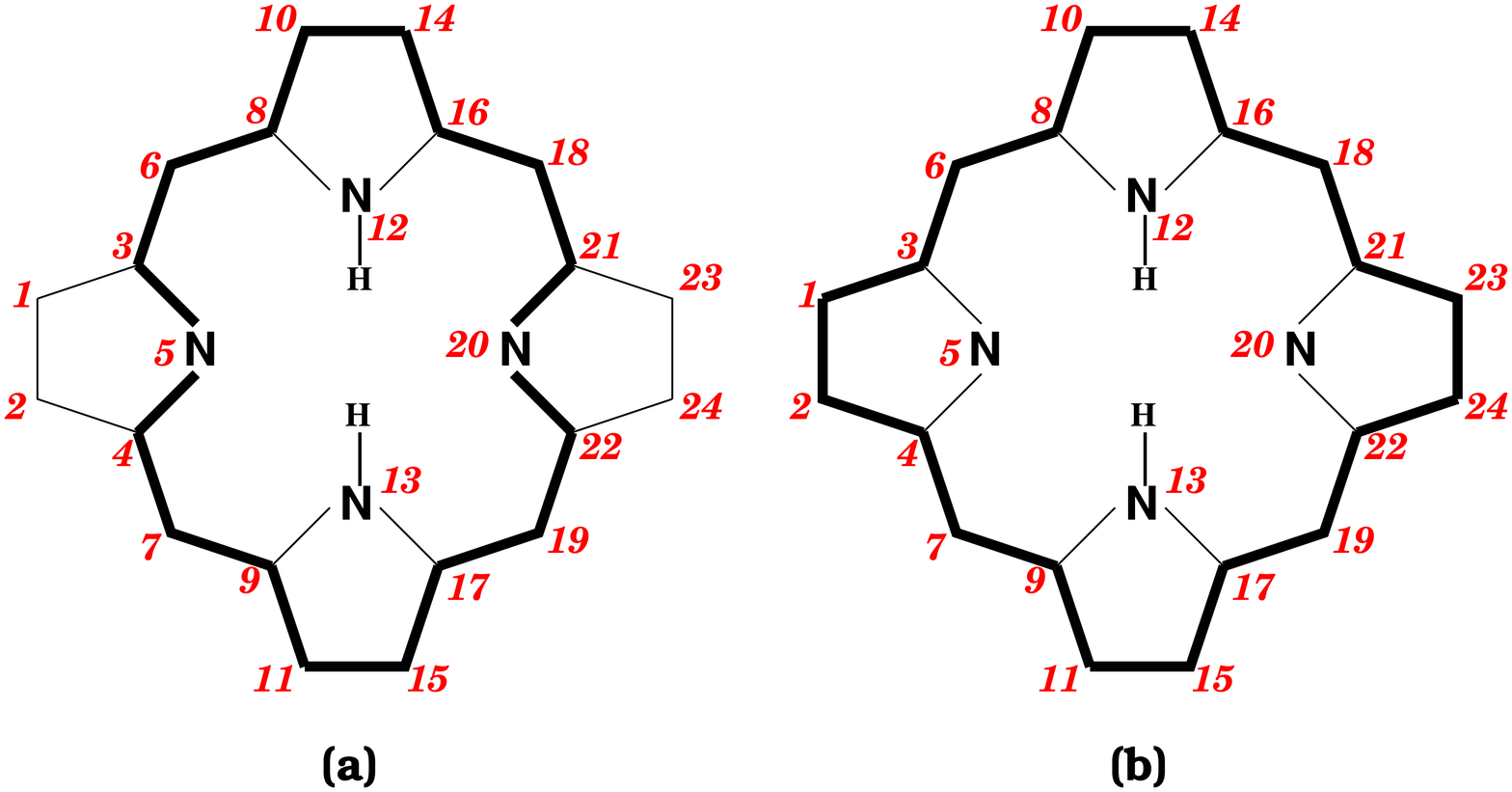}}
 \caption{Two possible equilibrium geometries in gs : (a) 18-sites annulenic 
structure and (b) 20-sites annulenic structure, shown by bold lines.}
\label{porfig4}
\end{center}
\end{figure}

It is known from the literature that the porphyrin can have two conjugation
pathways: an inner 18-annulene pathway and an outer 20-annulene pathway
(Fig. \ref{porfig4}). From our b.o. calculations we find that b.o.s are more
uniform ($0.55 \pm 0.03$) along the 18 annulene pathway.  
We should expect porphine
to behave more as an  18-annulene system than a 20-annulene system.
The two $C_{\beta}-C_{\beta}$ bonds have much higher bond order
($0.81$) and hence the 20 annulene description of porphines can be discarded.
This structure is in confirmity with X-ray structure \cite{webb} as well as other
theoretical work by Weiss {\it et al.} \cite{weiss}. From Fig. \ref{porfig4}, we see
that  pyrrole $N$ is not strongly involved in $\pi$-conjugation,
${\rm C_{\beta}-C_{\beta}}$ b.o. is small and hence we expect bond length of
${ \rm C_{\beta}-C_{\beta}}$ in  pyrrole rings to be large. In aza rings, N is
involved in the $\pi-$conjugation path and a reverse trend in b.o. is observed, namely,
${\rm {C_{\beta}-C_{\beta}}}$ b.o., is large and we should expect a shorter
$\rm {C_{\beta}-C_{\beta}}$ bond. The ${\rm C_{\alpha}-C_{\beta}}$ bond length is
increased \cite{simps} in aza rings.

\begin{figure}
\begin{center}
\hspace*{0cm}{\includegraphics[width=8.0cm,height=8.0cm]{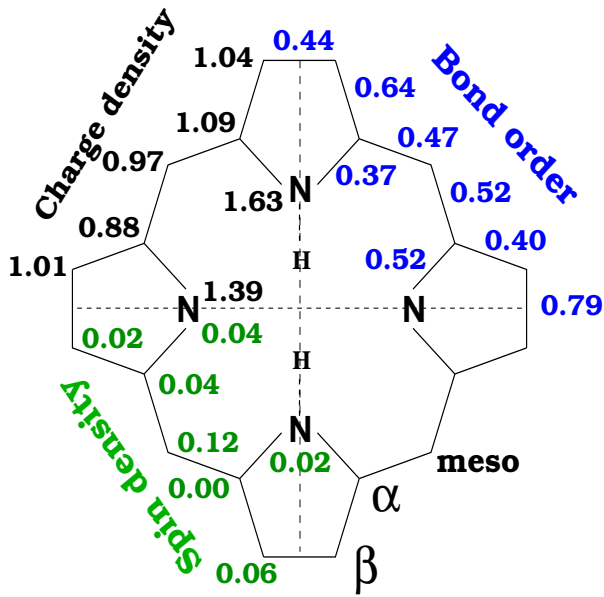}} \\
\end{center}
\caption{ Electron density, spin density and b.o.s for the lowest 
triplet state of porphine. Spin and charge densities are shown on the left part of the structure while b.o.s
are shown on the right part. The $\rm D_2$ symmetry gives value for other
half of the molecule.}
\label{porfig5}
\end{figure}

The devitation from the g.s.  bond order is given in Table \ref{portb5}.
In the first excited state, all the b.o.s become smaller compared to 
their values in gs geometry. The change in b.o. is only marginal 
in aza rings, while in pyrrole rings  ${\rm {C_{\beta}-C_{\beta}}}$ bonds become weaker. 
All the three ${\rm C-C}$ b.o.s are uniform. The $ {\rm C-N}$ bonds also 
become weaker. The $meso-$bonds connected to aza rings become 
weaker than those connected to pyrrole rings. Hence we expect the equilibrium 
geometry in the first excited state to have expansion of the whole $\pi$-system.   

In the second excited state, ${\rm C_{\beta}-C_{\beta}}$ bond of aza rings weakens
and ${\rm C_{\alpha}-C_{\beta}}$ bonds become stronger. Meso bonds connected to aza
rings become weaker and the rest become stronger compared to the
$1^{st}$ excited state. The pyrrole ring becomes more pyrrolic, i.e. the
${\rm C_{\beta}-C_{\beta}}$ b.o. reduces considerably and the 
${\rm C_{\alpha}-C_{\beta}}$ bond becomes a strong bond while the ${\rm C-N}$ bonds get weaker. 
This bond pattern is in confirmity with the charge distribution in the 
second excited state. The charge is depleted from the $N$ and accumulated on 
$C$ atoms. The b.o. pattern of third excited state is similar to that of second 
excited state.

In the fourth excited state, ${\rm C_{\beta}-C_{\beta}}$ bonds of aza rings become very
weak. The ${\rm C_{\alpha}-C_{\beta}}$ and the ${\rm C-N}$ bonds have almost the same
b.o. ($0.49\pm0.01$). This can again be attributed to the fact that
in the fourth excited state, there is charge accumulation on $\rm C_{\alpha}$ atoms 
and depletion at $N$ atom. The bond alternation of the $meso-$bonds
reduce. In pyrrole ring, all the three ${\rm C-C}$ bonds are almost uniform. Note
that in both the first excited and $4^{th}$ excited states, there is charge
accumulation at pyrrole $N$ and all the three ${\rm C-C}$ bonds have uniform bond
order.

In the fifth excited state, the aza b.o. and charge distribution pattern
are same as in fourth excited state. The charge density pattern at pyrrole ring
is similar to that of the third excited state and hence the b.o. pattern is
similar to that of third excited state, namely alternate single-double-single
bond for three ${\rm C-C}$ bonds and weaker ${\rm C-N}$ bonds.

In the sixth excited state, the charge depletion at aza $N$ is not significant
and the b.o. is slightly alternating for $\rm {C_{\alpha}-C_{\beta}}$ and 
$\rm {C-N}$ bonds.The ${\rm C_{\beta}-C_{\beta}}$ bond is more like a double bond. 
The pyrrole ring is less pyrrole like and the ${\rm C-N}$ bonds become weak.

In aza rings, when the charge is taken away from $N$, the b.o.s of 
${ \rm C_{\alpha}-C_{\beta}}$ bonds 
and ${\rm C-N}$ bonds would be uniform. On the other hand, when there is excess charge
at $N$, the bonds alternate. In pyrrole rings, when there is charge accumulation
at $N$, the three $C-C$ bonds become almost uniform and when the charge is taken
away from $N$, ${\rm C_{\alpha}-C_{\beta}}$ bonds become stronger and 
${\rm C_{\beta}-C_{\beta}}$ bonds become 
weaker. In all the excited states, either the aza ${\rm C-N}$ bonds have weak bond 
alternation or the pyrrole ${\rm C-N}$ bonds.

Spin densities, charge densities and b.o.s are shown in Fig. \ref{porfig5}, in the 
lowest triplet state of FBP. The electron density on $\rm C_{\beta}$ and 
$\rm C_{\alpha}$ carbon atoms are comparable to that in gs  while it decreases on the
nitrogen atom of aza-rings and remains constant for N atom on pyrrole-rings, compared 
to gs. At $C_{{\rm meso}}$ sites, electron density decreases in the lowest 
triplet state.  The values of $ {\rm C_{\beta}-C_{\beta}} $, ${\rm C_{\alpha} - 
C_{\beta}} $ and ${\rm C_{\alpha}-C_{{\rm meso}}}$ b.o.s in the aza-ring are almost 
same as in gs while $C_{\alpha}$-$C_{{\rm meso}}$ and ${\rm C_{\alpha}-C_{\beta}}$ 
b.o.s of pyrrole-ring increase while for ${\rm C_{\beta}-C_{\beta}}$ it decreases. 
$C-N$ b.o.s in aza-rings decrease but increase in pyrrole-rings. Spin density 
values on $C_{\beta}$ atoms in aza-ring are small but are relatively higher (0.045) 
in pyrrole-rings. Spin density values are very high on ${\rm C_{{\rm meso}}}$ sites
(0.16) and nitrogen atom (0.12) in the aza-rings while this value is very small on the 
nitrogen atom of pyrrole-rings. 

\subsection{Metallo-porphines}
Metallic porphines are biologically important molecules. There
are two categories of metallic porphines - regular and irregular porphines.
In the case of regular porphines, the metals have closed shells and in the
latter, they have partially filled shells. Our study concentrates on regular
metallic porphines. We have placed the metal ion at the center of the porphine
molecule and it acts only as a source of electric field, {\it i.e.}, we have not
taken the overlap of metallic $d-$ orbitals with porphine $\pi-$orbitals. 
The modified Hamiltonian to account for the effect of the metallic ion is
given by,
$\hat H=H_{PPP}+\sum q_M (\hat n_i-z_i) /r_{M_i}$
\noindent where $q_M$ is charge on the metal and $(\hat n_i-z_i)$ is 
effective charge operator at site $i$, $r_{M_i}$ is the distance between metal 
at the center and $i^{th}$ atom. 
In the presence of the metal ion, pyrrolic $N$ gets deprotonated and all the 
four five-membered rings become equivalent and hence the bond
 alternation observed in the neutral porphine molecule also vanishes.  
So, we have taken uniform transfer integrals for all the ${ \rm C-C}$ bonds and
a single value of $\epsilon_N$ and $U_N$ for all nitrogens. The uniform
 $\epsilon_N$ and $U_N$ used in the Hamiltonian are -7.8 eV and 12.34 eV respectively. 
We have varied the charge on the metal ion from 1 to 5 and studied the effect
of oxidation state of the metal ion on porphine spectra, ground 
state charge distribution and nonlinear optical properties. 
\subsubsection{ Ground state charge density and bond orders} 
Charge densities on porphine for different oxidation states of the 
central metal ion  are given in Table \ref{portb6}. Charge density on the 
$C_{\beta}$ atoms decreases while it increases on $C_{\alpha}$ atoms.
The charge density on $N$ atoms increases while those on $C_{{\rm meso}}$ 
atoms decrease. 
\begin{table}
\caption{Charge densities on unique sites of metallo-porphine for different oxidation 
states of the metal ion.}
~\\
\begin{center}
\begin{tabular}{|c|cccccc|} \hline
Site & \multicolumn{6}{c|}{Charge on metal site} \\ \cline{2-7}
 & $ 0$ & $1$ & $2$ & $3$ 
& $4$ & $5$  \\ \hline
   $C_{\beta}$ ~(aza) &  1.01 & 1.04   &     0.99   &      0.94   &    0.87    &    0.80 \\
     &       &    &             &              &             &           \\
   $C_{\alpha}$ ~(aza)    &  0.89 & 0.79   &     0.82   &      0.88   &    0.96    &    1.05  \\
     &       &    &             &              &             &           \\
   $N$ (aza)    & 1.41 & 1.73    &    1.82    &     1.88    &   1.92     &   1.95  \\ 
     &      &     &             &              &             &           \\
   $C_{meso}$     & 0.98 & 1.08    &    1.04    &     0.99    &   0.93     &   0.88  \\
     &      &     &             &              &             &           \\
   $C_{\alpha}$ ~(pyr)    & 1.08 & 0.78    &    0.82    &     0.89    &   0.97     &   1.06  \\
     &      &     &             &              &             &           \\
   $C_{\beta}$ ~(pyr)   & 1.02 & 1.05    &    0.99    &     0.92    &   0.85     &   0.77  \\
     &      &     &             &              &             &           \\
   $N$ ~(pyr)   & 1.63 & 1.79    &    1.86    &     1.91    &   1.94     &   1.96  \\ \hline
\end{tabular}
\label{portb6}
\end{center}
\end{table}
\begin{table}
\caption{ The variation in b.o.s of the porphine bonds with different oxidation states 
of the  central metal ion. B.o.s of bonds for which the value is not 
quoted  can be obtained by imposing $\rm D_2$ symmetry of the porphine.}
\begin{center}
~\\
 \begin{tabular}{|c|ccccc|} \hline
Bond & \multicolumn{5}{c|}{Charge on metal site} \\ \cline{2-6}
 &~~ $1$~~ &~~ $2$~~ &~~ $3$~~ &~~ $4$~~ &~~ $5$~~ \\ \hline
$C_{\beta}-C_{\beta}$     &  0.74   &  0.72   &  0.69   &  0.62   &  0.57  \\
      &         &             &           &           &   \\ 
$C_{\beta}-C_{\alpha}$     &  0.47   &  0.50   &  0.54   &  0.59   &  0.62  \\
$C_{\beta}-C_{\alpha}$     &  0.47   &  0.51   &  0.54   &  0.59   &  0.62  \\
      &         &             &           &           &   \\ 
$C_{\alpha}-N$     &  0.41   &  0.34   &  0.27   &  0.22   &  0.17  \\
$C_{\alpha}-N$     &  0.42   &  0.34   &  0.27   &  0.22   &  0.17  \\
      &         &             &           &           &   \\ 
$C_{\alpha}-C_{meso}$     &  0.59   &  0.60   &  0.60   &  0.59   &  0.57  \\
$C_{\alpha}-C_{meso}$     &  0.59   &  0.60   &  0.60   &  0.59   &  0.57  \\
      &         &             &           &           &   \\ 
$C_{\alpha}-C_{meso}$     &  0.57   &  0.58   &  0.59   &  0.59   &  0.60  \\
$C_{\alpha}-C_{meso}$     &  0.57   &  0.58   &  0.59   &  0.59   &  0.60  \\
      &         &             &           &           &   \\ 
$C_{\beta}-C_{\alpha}$    &  0.55   &  0.57   &  0.58   &  0.61   &  0.62  \\
$C_{\beta}-C_{\alpha}$    &  0.54   &  0.56   &  0.58   &  0.60   &  0.61  \\
      &         &             &           &           &   \\ 
                                                    
$C_{\alpha}-N$    &  0.35   &  0.28   &  0.22   &  0.18   &  0.14  \\
$C_{\alpha}-N$    &  0.33   &  0.27   &  0.22   &  0.18   &  0.14  \\
      &         &             &           &           &   \\ 
                                                    
$C_{\beta}-C_{\beta}$    &  0.61   &  0.61   &  0.59   &  0.56   &  0.52  \\ \hline
\end{tabular}
\end{center}
\label{portb7}
\end{table}

Bond orders in gs for metallo-porphine as a function of oxidation 
state on the metal ion are given in Table \ref{portb7}. 
As in the FBP molecule, ${\rm C_{\beta}-C_{\beta}}$ 
bonds in the aza rings are the strongest bonds except for $+5$ 
oxidation state. As the oxidation state of the metal ion increases, 
${\rm C_{\beta}-C_{\beta}}$ 
bond order decreases and the ${\rm C_{\beta}-C_{\alpha}}$ b.o.s increase. With the 
increase in $q_M$, as the charge on ${\rm N}$ reaches its maximum
value, the b.o. of all ${\rm C-N}$ bonds decrease sharply. 
Meso-bonds show minimum change; one set of b.o.s 
increases while the other set decreases, leading to small alternation 
in the $meso-$bonds. The asymmetry in the geometry of two rings 
reduces as the charge on the metal ion increases. The ${\rm C-N}$ bonds of 
aza rings have almost same b.o.s for all oxidation state while in pyrrole rings 
they are of same magnitude only for higher oxidation states. 

Our calculations are in good agreement with earlier semi empirical 
results. Poveda {\it et al.} \cite{poveda} found that ${\rm C_{\beta}-C_{\beta}}$
bonds are the strongest bonds and the ${\rm C_{\beta}-C_{\alpha}}$ bonds are the weakest
bonds in both Mg Octaethyl tetra phenyl porphyrin (OETPP) and $ZnOETPP$. For charge 
$+2$ we also find 
that the ${\rm C_{\beta}-C_{\beta}}$  bonds are the strongest bonds and 
${\rm C_{meso}-C_{\alpha}}$ bonds are stronger than ${\rm C_{\alpha}-C_{\beta}}$ bonds. 
The ${\rm C-N}$ bonds become weaker in the presence of the metal ion. 
All the ${\rm C-N}$ b.o.s of FBP are always higher than those in the metallo-porphine. 
Pyrrolic ${\rm C-N}$ bonds have a slight alternation in bond order 
pattern at lower oxidation states. FBP in the excited states show 
alternate b.o.s for ${\rm C-N}$ bonds either in the aza rings or in the 
pyrrole rings.  While FBP has a 18-annulenic structure, metallo-porphine shows 
20-annulene like structure when the charge on the metal ion increases.

\subsubsection{Optical properties}
   The metallo-porphines have higher symmetry than FBP; 
former has $D_{4h}$ point group symmetry while the point group symmetry 
 of the latter is $D_{2h}$. Hence the spectra of metallo-porphines 
have more degenerate excitations than in FBP. 

The characteristic optical spectra of metallo-porphines consists of a $Q$-band lying 
between 500-600 nm. $Q$-bands have two  nearly degenerate levels separated by 1250 
$cm^{-1}$. It was originally identified as a vibronic progression and assigned 
$Q_{0->0}$ and $Q_{0->1}$ transition \cite{platt}. $B$-band which is the most 
intense band lies between 380 and 420 nm. Other $L$, $M$ and $N$ bands are 
comparatively weak and appear between 325 nm and 215 nm. 

\begin{table}
\caption{Excitation gaps for singlet-singlet transition  of five lowest
optically allowed states and transition dipole moments for different metallic
charges. }
~\\
\begin{center}
\begin{tabular}{|c|ccc|} \hline
$q_M$ & Excited & Optical gap & ${|\mu |}$   \\ \cline{2-4}
  & state &  (in eV) &  (in a.u.) \\ \hline
   &  1B  &  1.70 &  0.80  \\
   &  2B  &  2.04 &  0.08  \\
1  &  3B  &  3.94 &  0.66   \\
   &  4B  &  4.11 &  2.30  \\
   &  5B  &  5.37 &  0.26  \\  
   &      &       &  \\
   &  1B  &  1.69  & 0.46 \\
   &  2B  &  1.97  & 0.16  \\
2  &  3B  &  4.00  & 2.43 \\
   &  4B  &  4.22  & 0.66  \\
   &  5B  &  5.36  & 0.59   \\ 
   &      &       &  \\

   &  1B  &  1.70  & 0.06 \\          
   &  2B  &  1.84  & 0.41 \\          
3  &  3B  &  3.87  & 2.53 \\          
   &  4B  &  4.23  & 0.58 \\          
   &  5B  &  5.13  & 0.55 \\  

   &      &       &  \\
   &  1B  &  1.66  & 0.66 \\
   &  2B  &  1.73  & 0.71  \\
4  &  3B  &  3.59  & 2.41 \\
   &  4B  &  4.36  & 1.38  \\
   &  5B  &  5.20  & 0.40  \\  

   &      &       &  \\
   &  1B  &  1.65  & 1.02  \\
   &  2B  &  1.68  & 1.06 \\
5  &  3B  &  3.62  & 2.26 \\
   &  4B  &  4.31  & 1.32  \\
   &  5B  &  4.78  & 0.10 \\  \hline

\end{tabular}
\end{center}
\label{portb8}
\end{table}

The five lowest excitation gaps along with transition dipole
moments are given in Table \ref{portb8} for five different oxidation 
states of the metal ion. The first allowed state lies about $ \approx 1.70$ eV
above the gs. For $q_M=+1$ and $q_M=+2$ transition dipoles decrease with 
increasing charge. The lowest level has almost no intensity
 for dipole transition from gs for $q_M=+3$ and the second level has significant 
intensity. We find that for metal ion oxidation states upto $q_M=+3$, 
 $Q$-bands are well separated. But with increase in $q_M$, the separation 
between $Q$-bands decreases. For $q_M>+3$ on the metal ion, this separation 
is very small, and both states have large transition dipoles to the gs 
although with different polarizations. The intensity for 
second $Q$-band transition for $q_M=+2$ is weak and is consistent 
with experimental results \cite{gouter63}. Experiments show that the 
first excited state of metallo-porphines is slightly lower in energy 
than in FBP, irrespective of substitutions at the $meso-$positions as shown 
in reference \cite{jac89}.  There is only a small variation (maximum of 0.1eV) in 
optical gap for different substituents. 

From our DMRG results, the strongly optically allowed band, $B$-band, 
lies between $3.8 \pm 0.2 eV$ above the gs for all  values of $q_M$. These levels 
show red shift of about $0.2 eV$ as $q_M$ is increased from $+1$ to $+5$. 
For $q_M=+2$, the experimental gaps are about $3.22 eV$ for Zn and $3.18 eV$ 
for Mg(etio)Porphine \cite{gouter63}. Time dependent density functional theory 
calculation of Minaev {\it et al.} has a blue shift of $~0.23$ \cite{minaagren}, 
while symmetry adapted cluster CI study of Hasegawa {\it et al.} has a blue shift of
 $~0.5$ eV for the $B$ band compared to the experimental results \cite{Hasegawa}. 
The blue shift in our DMRG studies is 0.6 eV. 
Complete active space second order perturbation theory (CASPT2) 
calculation of Rubio {\it et al.} underestimates the experimental excitation 
energies by $0.3$ to $0.5 eV$ \cite{rubio}. 

We observe a transition of medium intensity around $4.2\pm0.1$ eV in all 
metallo-porphines, corresponding to experimental absorption peaks for $N$-bands of 
Mg(etio)Porphine and TPPMgP observed at $3.85eV$ and $3.97eV$ respectively. 
Our result overestimates this gap by  $ \approx 0.3 eV$. There is also a weak 
$L$-band around $5.3 \pm 0.1 eV$. To the best of our knowledge, there are no
experimental results for system with $q_M\ge 3$, for regular metalloporphines 
for comparison with DMRG results. 
\begin{table}
\caption{ Triplet state energies, T-T gaps and corresponding transition dipole 
moments of metallo-porphine for different oxidation state of metal ions.}
~\\
\begin{center}
\begin{tabular}{|ccccc|} \hline
$q_M$ & state& Triplet  & T-T gap & $|\mu|$  \\ 
  &    index & energies (eV) &(eV)&  (in a.u.) \\ \hline
  &1 & 3.31& 1.29&  0.30\\
1 &2 & 3.48& 1.46&  0.59\\
  &3 & 4.11& 2.09& 0.28 \\
  &4 & 4.91& 2.89& 0.17 \\ 

   &   &   &   &  \\
   &1 &  3.01&1.27& 0.34\\
2  &2 & 3.33 &1.59&  0.29 \\
   &3 & 3.84& 2.10&  0.25 \\
   &4 & 4.41& 2.67& 0.19 \\ 
   &   &   &   &  \\

   &1 & 2.99& 1.27 &  0.19 \\
 3 &2 & 3.31 &1.59 & 0.08 \\
   &3 & 3.62 &1.90 &  0.47 \\
   &4 & 4.23 & 2.51 & 0.12 \\ 
   &   &   &   &  \\

   &1 &3.29 &1.69 & 0.07 \\
 4 &2 &3.45 &1.85 & 0.09 \\
   &3 &3.99 &2.39 & 0.48 \\
   &4 &4.11 &2.51 & 0.24 \\ 
   &   &   &   &  \\

   &1 &2.76 &1.60 & 0.69 \\
 5 &2 &3.25 &2.09 &0.51 \\
   &3 &3.30 &2.14 & 0.55\\
   &4 &3.73 &2.57 & 0.26\\  \hline
\end{tabular}
\end{center}
\label{portb9}
\end{table}
\subsubsection{Triplet-Triplet spectra}
 In Table \ref{portb9}, triplet energies and triplet-triplet 
(T-T) gaps and corresponding transition dipole moments are given. Goutermann {\it et  al.} 
\cite{gouterkhalil} observed a weak T-T absorption at $1.41eV$ and 
a high energy T-T absorption at $1.58eV$ with high 
intensity in case of TPPZn. Sapunov {\it et al.} observed a peak at $1.53 eV $ 
for the same system in dimethyl phthalate solvent \cite{sapunov} while Pileni {\it et al.} observed 
peaks at $ 1.48 eV$ and $1.68 eV$ in DODAC solution of TPPZn. 
Goutermann {\it et al.} also measured the phosphorescence of $(TPP)SnCl_2$,
which exhibits similar behavior as TPPZn. DMRG results for $q_M=2$ have 
the first T-T transition at $1.27 eV $ and the second at $1.59 eV$. 
The first transition is red-shifted compared to the experimental value. In our calculation we observe
 that the intensity of the first peak is comparable to the second peak. Other
 higher excited states have relatively weak intensities. 

   As the oxidation state of the metal ion increases, the first T-T gap remains almost a 
constant at $1.27eV$ up to $q_M=3$ and then it increases for charge $q_M=4$ and 5. 
All the T-T gaps for charge $q_M=+2$ and $q_M=+3$ are similar. For 
charge $q_M=4$, the first observable transition is at $2.39 eV$. For $q_M=5$, 
all the T-T transitions have significant transition dipoles. 

\section{Nonlinear optical properties of porphines}
Hitherto, the dynamical NLO response of porphines 
has been studied using a sum-over-state (SOS) method in  
conjunction with restricted CI studies. 
As has been stated before, DMRG method is far superior to other 
many-body technique for quasi-one-dimensional systems. Besides, 
the DMRG method has been combined with the CV method to bypass 
computing only a part of the excited state spectrum as required in 
the SOS method. 
The CV method incorporates the full excitation spectrum in the 
truncated DMRG basis, which in our scheme corresponds to about 
45,000 excited states for the DMRG cut-off of $m=210$. 

We have calculated the linear and third order nonlinear optic 
coefficients corresponding to third harmonic generation (THG). These r
esponse coefficients have been calculated at three different frequencies. 
These computational results have been obtained for both FBP and regular 
metalloporphines.

\subsection{Results and Discussion}

\begin{figure}[h]
\begin{center}
{\includegraphics[width=9.0cm,height=9.0cm]{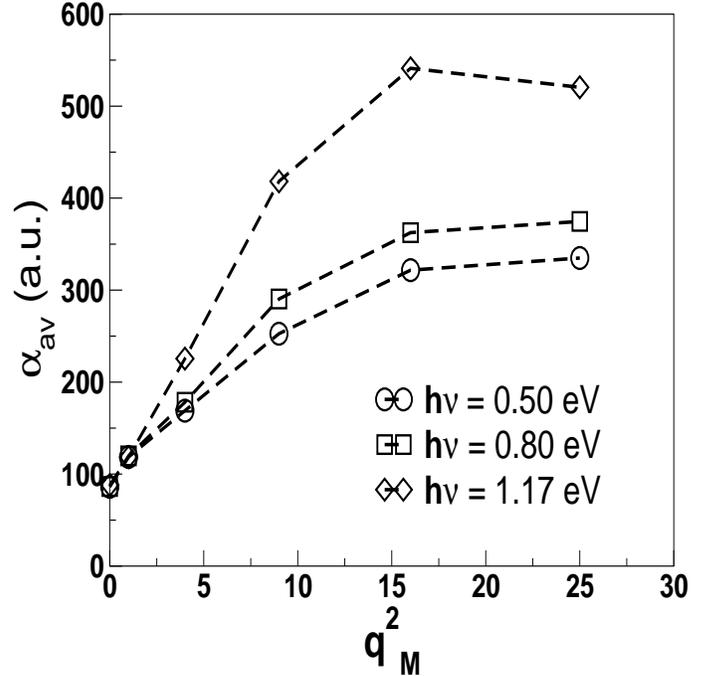}} \\
\caption{ Tumbling averaged  linear polarizability $(\alpha)$ vs $q_m$ at
three different frequencies.}
\label{porfig6}
\end{center}
\end{figure}
\begin{table}
\begin{center}
\caption{ Tumbling averaged THG coefficient $\gamma_{av}$ in $10^3$ a.u, for different 
excitation frequencies and oxidation states of the central metal ion.}
~\\
\begin{tabular}{|c c c c|} \hline
~~ $q_M$ ~~ & ~~ $\omega$ ~~ & ~~Real ($\gamma$)~~ &~~ Img($\gamma$) \\ \hline
 & 0.65 & 158.20 & 0.00 \\
0.0 & 0.80 & 169.51 & 0.00 \\
 & 1.17 & 220.14 & 0.01\\ 

 &    &   &  \\

 & 0.50 & 1584.88& 0.93 \\
1.0 & 0.80 & 1276.63& 0.04 \\
 & 1.17 & 1367.07& 1.05 \\ 

 &    &   &  \\
 & 0.65 & 1237.32&0.48 \\
2.0 & 0.8  & 1122.09&0.20\\
 & 1.17 & 1076.71&0.38 \\ 

 &    &   &  \\
 & 0.65 &1599.05&1.08 \\
3.0 & 0.80 &1220.09&0.30\\
 & 1.17 &1117.39&0.13\\ 

 &    &   &  \\
 & 0.50 & 2148.68&1.32 \\
4.0 & 0.80 & 1711.20&0.18\\
 & 1.17 & 1756.68&0.88 \\ 

 &    &   &  \\
 & 0.65 &3171.83&1.60 \\
5.0 & 0.80 &2719.45&0.32 \\
 & 1.17 &3531.41&4.75 \\\hline

\end{tabular}
\label{gammat}
\end{center}
\end{table}

For small system sizes (up to about 12 sites), the DMRG calculations can be 
compared with exact results. This has been done for polarizability 
$\alpha$, as well as for the THG coefficients ($\gamma$) for various
components and at different frequencies. The accuracy is typically better 
than $1\%$ for dominant components away from resonances. 

 As shown in the Fig. \ref{porfig6}, tumbling averaged linear polarizability 
$\alpha_{av}$ increases quadratically with increasing oxidation state, $q_M$, on 
the metal ion at all the frequencies we have studied. However, there 
seems to be saturation in the $\alpha_{av}$ value for $q_M>3$, at all frequencies.
This trend seems to be consistent with the optical gaps and associated transition 
dipole moments for the low-lying states for different $q_M$ values, although 
quantitative inference about $\alpha_{av}$ cannot be arrived at based on the 
few low-lying state properties we have computed.

The tumbling averaged THG coefficients ($\gamma_{av}$), are presented in 
Table \ref{gammat}. The THG coefficients, at all the three frequencies we have
computed are about one order of magnitude smaller for FBP than regular 
metallo-porphines. Increase in $q_M$ leads to increase in $\gamma_{av}$, away 
from resonances. 
$\gamma_{av}$ value for $q_M=1$, 2 and 3 are very similar and much larger 
than the FBP values. For $q_M>3 $, there is a large increase in $\gamma_{av}$, 
with $q_M$. Thus metallo-porphines with large oxidation state on the metal ion 
are expected to exhibit large THG responses.

\section{Conclusion}
In Summary, we have employed the symmetrized DMRG method for studying 
the linear and nonlinear optical properties of FBP and metallo-porphines within an 
interacting $\pi$-electron model. We note that the computed singlet and triplet 
low-lying excited state energies and their charge densities are in excellent agreement 
with experimental as well as many other theoretical results. From our bond order 
calculation, we conclude that porphine has 18-annulenic structure in the ground 
state and the molecule gets
expanded upon excitation. We have modelled the metalloporphine  by taking
an effective electric field due to the metal ion and computed the excitation spectrum.
Metalloporphines have $D_{4h}$ symmetry and hence more degenerate excited states.
The ground state of Metalloporphines show 20-annulenic structure, as the charge
on the metal ion increases. The linear polarizability seems to increase with the
charge initially and then saturates. The same trend is observed in third order
polarizability coefficients.

{\bf Acknowledgments. } MK thanks UGC India for financial support.  
SR thanks DST India for funding through different programmes.

\end{document}